\documentclass[prl,superscriptaddress,amsmath,amssymb,twocolumn,nofootinbib]{revtex4-2}

\usepackage{graphicx,bm,amsmath}
\usepackage[usenames,dvipsnames]{xcolor}
\usepackage[colorlinks,citecolor=NavyBlue,linkcolor=NavyBlue,urlcolor=NavyBlue,]{hyperref}

\usepackage[bbgreekl]{mathbbol}
\usepackage{xcolor}
\usepackage[normalem]{ulem}
\usepackage{algorithm}
\usepackage{braket}
\usepackage{algpseudocode}
\usepackage{comment}
\usepackage{amsthm}
\usepackage{enumerate}
\usepackage{orcidlink}

\newcommand{\be}{\begin{equation}}
\newcommand{\ee}{\end{equation}}
\newcommand{\bec}{\begin{equation*}}
\newcommand{\eec}{\end{equation*}}
\newcommand{\bea}{\begin{eqnarray}}
\newcommand{\eea}{\end{eqnarray}}

\newcommand{\titleinfo}{
Universal Spreading of Nonstabilizerness and Quantum Transport}

\newcommand{\Tr}{\text{Tr}}   
 
\begin{document}
\title{\titleinfo}

\author{Emanuele Tirrito~\orcidlink{0000-0001-7067-1203}}
\email{etirrito@ictp.it}
\affiliation{The Abdus Salam International Centre for Theoretical Physics (ICTP), Strada Costiera 11, 34151 Trieste, Italy}
\affiliation{Dipartimento di Fisica ``E. Pancini", Universit\`a di Napoli ``Federico II'', Monte S. Angelo, 80126 Napoli, Italy}

\author{ 
Poetri Sonya Tarabunga~\orcidlink{0000-0001-8079-9040}}
\email{poetri.tarabunga@tum.de}
\affiliation{Technical University of Munich, TUM School of Natural Sciences,
Physics Department, 85748 Garching, Germany}
\affiliation{Munich Center for Quantum Science and Technology (MCQST),
Schellingstr. 4, 80799 M\"unchen, Germany}

\author{Devendra Singh Bhakuni~\orcidlink{0000-0003-3603-183X}}
\affiliation{The Abdus Salam International Centre for Theoretical Physics (ICTP), Strada Costiera 11, 34151 Trieste, Italy}
\email{dbhakuni@ictp.it}
\author{Marcello Dalmonte~\orcidlink{0000-0001-5338-4181}}
\affiliation{The Abdus Salam International Centre for Theoretical Physics (ICTP), Strada Costiera 11, 34151 Trieste, Italy}

\author{Piotr Sierant~\orcidlink{0000-0001-9219-7274}}
\email{psierant@icfo.net}
\affiliation{Barcelona Supercomputing Center, Barcelona 08034, Spain}

\author{ 
Xhek Turkeshi~\orcidlink{0000-0003-1093-3771}}
\email{turkeshi@thp.uni-koeln.de}
\affiliation{Institut f\"ur Theoretische Physik, Universit\"at zu K\"oln, Z\"ulpicher Strasse 77, 50937 K\"oln, Germany}

\begin{abstract}
We investigate how transport properties of $\mathrm{U}(1)$-conserving dynamics impact the growth of quantum resources characterizing the complexity of many-body states. 
We quantify wave-function delocalization using participation entropy (PE)—a measure rooted in the coherence theory of pure states—and assess nonstabilizerness through stabilizer Rényi entropy (SRE).
Focusing on the XXZ spin chain initialized in domain-wall state, we demonstrate universal power-law growth of both PE and SRE, with scaling exponents explicitly reflecting the underlying transport regimes—ballistic, diffusive, or KPZ-type superdiffusive.
Our results establish a solid connection between quantum resources and transport, providing insights into the dynamics of complexity within symmetry-constrained quantum systems.
\end{abstract}

\maketitle
\paragraph{Introduction.}
Transport phenomena in many-body systems are traditionally understood as emergent classical behavior arising from underlying quantum dynamics~\cite{bertini2015macroscopic,bertini2021finite, Landi22}. 
When local conservation laws are present, the late-time evolution of the system is well captured by classical hydrodynamics, which describes how conserved quantities diffuse or propagate through the system~\cite{doyon2025generalized}.  
This effective description typically applies regardless of whether the microscopic dynamics is quantum or classical, leading to the widespread expectation that transport exponents and fluctuations admit a classical interpretation. 
Indeed, in chaotic systems, coarse-grained observables such as currents and densities exhibit self-averaging behavior and are governed either by classical diffusion laws or by the more intricate framework of fluctuating hydrodynamics~\cite{Crossley2017, Michailidis24Corrections, Wienand2024}.
As a result, transport is commonly regarded as a classical emergent phenomenon, largely insensitive to genuinely quantum features of the evolving state.

\begin{figure}[t!]
   \centering
\includegraphics[width=\linewidth]{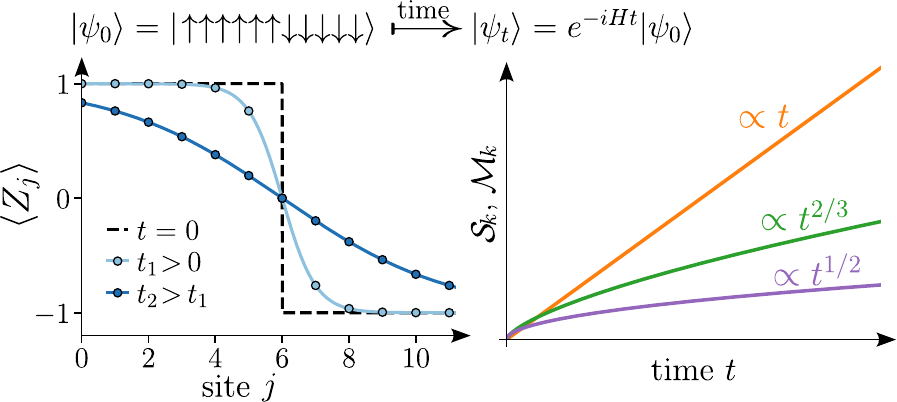}
\caption{ 
\textbf{Sketch of the investigated setup:} An \( N \)-qubit system is prepared in a product state \( \ket{\psi_0} \) in a domain-wall configuration. Coherent evolution under a \( U(1) \)-symmetric Hamiltonian \( H \) leads to the melting of the domain wall and the restoration of a uniform magnetization profile, represented by \( \braket{Z_j} = \braket{\psi_t|Z_j|\psi_t} \), where \( Z_j \) is proportional to the \( z \)-component of the spin-1/2 operator. The initial state has vanishing participation entropy, \( \mathcal{S}_k(\ket{\psi_0}) = 0 \), and stabilizer Rényi entropy, \( \mathcal{M}_k(\ket{\psi_0}) = 0 \). Time evolution under \( H \) increases these resources as \( \mathcal{S}_k, \mathcal{M}_k \propto t^{1/z} \), where \( z \) is the dynamical exponent characterizing transport in the system: \( z = 2 \) for diffusion, \( z = 3/2 \) for superdiffusion, and \( z = 1 \) for ballistic transport.}
\label{fig:sketch_transport}
\end{figure}

Despite its classical appearance, transport originates from the unitary dynamics of an underlying quantum state. This raises a fundamental question: to what extent do genuinely quantum features—those not reducible to classical probability distributions—participate in or influence transport? 
While entanglement is a widely used probe of quantum dynamics, it primarily captures local correlations between subsystems~\cite{ljubotina2017spin, Misguich17}. In contrast, global features such as quantum coherence and nonstabilizerness (or magic resources) are extensive properties of the wavefunction that characterize more intricate forms of quantum complexity and directly impact classical simulability. Coherence~\cite{baumgratz2014quantifying, Levi14} relates to anticoncentration~\cite{hangleiter2018anticoncentration,Bouland19,boixo2018characterizing,dalzell2022random, Turkeshi24Delocalization,lami2025anticoncentration,sauliere2025universalityanticoncentrationchaoticquantum,magni2025anticoncentrationcliffordcircuitsbeyond}, while nonstabilizerness quantifies deviations from stabilizer structures~\cite{bravyi2005universalquantumcomputation, knill2005quantum}. Both serve to distinguish dynamically generated states as classically tractable or intrinsically complex~\cite{aaronson2011computational, hangleiter2023computational}. 
This perspective leads us to ask: \textit{How do conservation laws constrain the spreading of these resources? Do coherence and nonstabilizerness evolve in tandem with transport, and can they reveal universality classes beyond the reach of entanglement-based diagnostics?  }

To address these questions, we study $\mathrm{U}(1)$-symmetric Hamiltonian dynamics in the XXZ spin chain, which features ballistic, diffusive, and superdiffusive regimes depending on anisotropy~\cite{PhysRevB.58.R2921,PhysRevLett.122.210602,PhysRevLett.117.207201,PhysRevB.96.115124,PhysRevB.97.045407,ljubotina2017spin,ljubotina2019ballistic,rosenberg2024dynamics,denardis2023nonlinear,deNardis2019Anomalous,deNardis2020Superdiffusion} as well as the disordered XXZ model which exhibits diffusive transport for small disorders strengths~\cite{PhysRevLett.117.040601}. 
We track quantum complexity growth using the participation entropy (for coherence)-- for which we introduce a scalable algorithm -- and the stabilizer Rényi entropy (for nonstabilizerness), and show that these a priori distinct measures are tightly correlated in the presence of $\mathrm{U}(1)$ symmetry. As shown in Fig.~\ref{fig:sketch_transport}, starting from a product state, the system builds up coherence and magic resources under time evolution, with both entropies growing as power laws governed by the transport exponent $z$: ballistic ($z = 1$), diffusive ($z = 2$), and KPZ ($z = 3/2$) up to the time-scales considered. These universal scaling laws reveal that quantum complexity is not merely compatible with hydrodynamic behavior but actively encodes it. 
Unlike entanglement entropy, which fails to discriminate between these regimes, participation and stabilizer entropies provide sensitive and complementary probes of dynamical complexity. 
In summary, our results demonstrate that quantum resource measures such as nonstabilizerness and anticoncentration are robust and informative diagnostics of transport in $\mathrm{U}(1)$-symmetric systems. They go beyond entanglement in 
capturing universal features of complexity growth, even in interacting Hamiltonians, and open new directions for understanding thermalization in symmetry-constrained quantum matter.

\paragraph{Preliminaries.} We consider a one-dimensional chain of $L$ qubits with Hilbert space $\mathcal{H} = (\mathbb{C}^2)^{\otimes L}$ and total dimension $D = 2^L$. The Pauli group $\mathcal{P}_L$ consists of all tensor products of Pauli operators with phases $\pm1$, $\pm i$
\begin{equation}
\tilde{\mathcal{P}}_L = \left\lbrace \omega P_{1} \otimes \cdots \otimes P_{L} \,\middle|\, P_j\in \{I,X,Y,Z\},\omega\in\{\pm1,\pm i\} \right\rbrace.
\end{equation}
The unsigned Pauli group $\mathcal{P}_L =\tilde{\mathcal{P}}_L /\mathrm{U}(1)$ identifies the strings modulo their phases. 
The Clifford group $\mathcal{C}_L \subset \mathrm{U}(D)$ is the normalizer of $\mathcal{P}_L$, consisting of all unitaries that conjugate Pauli operators to Pauli operators. It is generated by $S = \sqrt{Z}$, $H = (X + Z)/\sqrt{2}$, and CNOT gates, and forms a unitary 3-design for qubits, approximating Haar randomness up to third moments~\cite{Web16, Zhu17}. A pure state is called a \emph{stabilizer state} if it can be prepared by Clifford operations acting on $\ket{0}^{\otimes L}$.

In this work, we focus on systems with a $\mathrm{U}(1)$ symmetry associated with conservation of total magnetization, $S^z = \sum_{j=1}^L Z_j$. The corresponding $\mathrm{U}(1)$-symmetric Clifford group is defined as
\begin{equation} \label{eq:u1_clifford}
    \mathcal{C}_L^{\mathrm{U}(1)} = \left\lbrace U \in \mathcal{C}_L \,\middle|\, [U, S^z] = 0 \right\rbrace.
\end{equation}
This constraint significantly reduces the set of allowed Clifford gates, forbidding operations that mix magnetization sectors. The $U(1)$-symmetric Clifford operations are composed of qubit permutations, together with CZ and S gates. As a result, the group $\mathcal{C}_L^{\mathrm{U}(1)}$ forms a symmetric unitary 1-design but not a 2-design~\cite{PRXQuantum.4.040331, PhysRevResearch.6.043292, PhysRevLett.134.180404}, approximating the $\mathrm{U}(1)$ Haar-random unitaries only up to the first moment, and hence with limited capacity for scrambling of quantum information. 

The symmetry also restricts the set of states that can be prepared with U(1)-symmetric Clifford operations: in any magnetization sector, only computational basis states can be prepared from computational basis state input $\bigotimes_{i=1}^L \ket{x_i}$ by acting with $\mathcal{C}_L^{\mathrm{U}(1)}$~\cite{PhysRevResearch.6.043292}. States such as $\ket{\psi_{0} }=(\ket{01} + \ket{10})/\sqrt{2}$, although lying in a fixed magnetization sector, cannot be prepared by symmetry-respecting Clifford circuits and thus require magic resources from the perspective of the $\mathcal{C}_L^{\mathrm{U}(1)}$ group. 

\begin{figure*}[t]
   \centering
\includegraphics[width=1.0\linewidth]{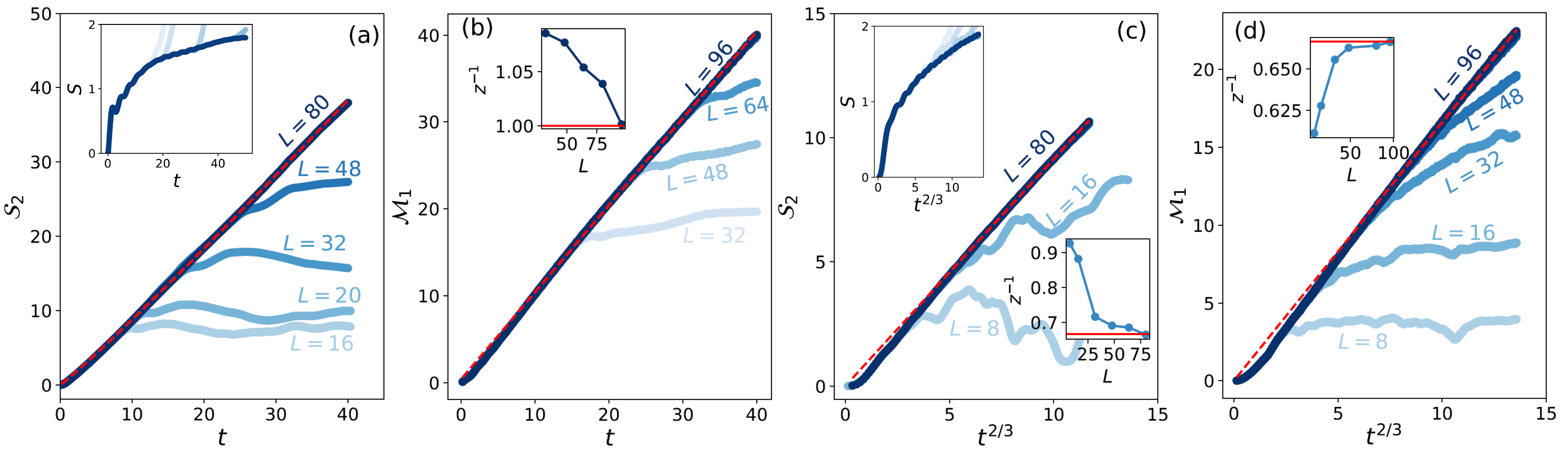}
\caption{\textbf{Ballistic and superdiffusive regime:} Time evolution of (a) PE and (b) SRE for $J_z=0.25$ and different system sizes $L$. Both PE and SRE exhibit a linear increase in time, consistent with ballistic transport dynamics characterized by a dynamical exponent $z=1$. 
In (c) and (d), we show PE and SRE for $J_z=1$ and various system sizes $L$. Both measures exhibit power-law growth consistent with superdiffusive spin transport characterized by a KPZ scaling exponent $z \sim 3/2$. The insets in (b,d) shows finite-size scaling analysis of the extracted dynamical exponent $z$, confirming ballistic and robust KPZ universality, while in (a,c), we present the dynamics of the entanglement entropy.
}
\label{fig:PE_SRE_ballistic_XXZ}
\end{figure*}

\paragraph{Stabilizer and participation entropy.} 
The resource theory~\cite{chitambar2019quantumresourcetheories} in which Clifford operations $\mathcal{C}_L$ are treated as free operations is known as the theory of nonstabilizerness~\cite{Veitch2014theresourcetheory, liu2022manybody, haug2023stabilizerentropiesand, tarabunga2024nonstabilizerness}.
As a measure of nonstabilizerness, we employ the stabilizer Rényi entropy (SRE)~\cite{leone2022stabilizerrenyientropy}, defined for a pure state $\ket{\psi}$ as
\begin{equation} \label{eq:sre_def}
\mathcal{M}_{k}(|\psi\rangle) = \frac{1}{1-k} \log \left[ \sum_{P \in \mathcal{P}_L} \frac{|\langle \psi | P | \psi \rangle|^{2k}}{D} \right],
\end{equation}
where $\mathcal{P}_L$ is the Pauli group (modulo phases) and $D = 2^L$, and $\mathcal{M}_{1}$ is defined by the limit $k \rightarrow 1$. The SRE satisfies the essential properties of a measure of nonstabilizerness: (i) it is faithful, i.e., (ii) it vanishes if and only if $\ket{\psi}$ is a stabilizer state; it is invariant under the action of Clifford unitaries; (iii) it is additive on product states~\cite{haug2023stabilizerentropiesand,gross2021schurweylduality}. The possibility of efficient numerical estimation of SRE has enabled widespread applications in many-body settings~\cite{haug2023quantifyingnonstabilizernessof,tarabunga2024mps,Tarabunga2024magicingeneralized,tirrito2024quantifying,tarabunga2023manybodymagic}, including studies of chaos~\cite{leone2021quantum, Turkeshi25Pauli}, localization~\cite{falcao2025magic}, and criticality~\cite{haug2023quantifyingnonstabilizernessof,FalcaoU1,PhysRevB.103.075145, tarabunga2024critical,tarabunga2025efficientmutualmagicmagic,oliviero2022magic,ding2025evaluating}. Moreover, notable progress has been made in their experimental measurements~\cite{Oliviero2022measuring,niroula2023phase,bluvstein2024logical,haug2024efficient,haug2025efficientwitnessingtestingmagic}.

Participation entropy (PE) characterizes the spread of wave-function in the computational basis, and is defined as
\begin{equation} \label{eq:pe_entropy}
\mathcal{S}_k(|\psi\rangle) = \frac{1}{1-k} \log \left[ \sum_{x=0}^{D-1} |\langle x | \psi \rangle|^{2k} \right],
\end{equation}
where $k$ is the Rényi index, and $\mathcal{B}=\{|x\rangle\;|\;x=0,\dots,D-1\}$ is the computational basis. The PE is a well-established measure of coherence of pure states~\cite{baumgratz2014quantifying}, and has been used in studies of many-body ground states and phase transitions~\cite{Stephan09, Stephan10, Zaletel11, Alcaraz13, laflorencie2014spin, luitz2014improving, luitz2014shannon,  Lindinger19, Pausch21}, systems out-of-equilibrium~\cite{backer2019multifractal, mace2019multifractal, sierant2022universal, Turkeshi24Delocalization} and can be measured experimentally~\cite{Liu25}. The U(1)-symmetric Clifford group acts diagonally in the computational basis. Hence, PE remains invariant under the action of $\mathcal{C}_L^{\mathrm{U}(1)}$ and can be perceived as a measure of magic resources from the perspective of $\mathcal{C}_L^{\mathrm{U}(1)}$ group, being strictly positive when the state becomes a non-trivial superposition of the computational basis states.

The SRE and the PE correspond to distinct resource theories, and are sensitive to different aspects of the complexity of the many-body state. However, in the following, we argue that under U(1)-symmetric dynamics, the time evolution of PE and SRE share similar features, which may reflect the transport properties of the system.

In $\mathrm{U}(1)$-symmetric systems, the SRE becomes dominated by diagonal Pauli strings, aligning it structurally with the PE. Eqs.~\eqref{eq:sre_def} and~\eqref{eq:pe_entropy} thus both describe basis-dependent spreading—one in operator space, the other in Hilbert space~\cite{haug2023quantifyingnonstabilizernessof,turkeshi2023measuring,collura2024quantummagicfermionicgaussian,haug2024efficient,turkeshi2024magic,tarabunga2024mps,magni2025anticoncentrationcliffordcircuitsbeyond}. This motivates a direct comparison~\cite{tarabunga2024nonstabilizerness}. In the following, we derive explicit analytical bounds linking these two quantities, establishing a formal connection between nonstabiilzerness and anticoncentration for U(1)-symmetric states. 
Our first inequality reads
\begin{equation} \label{eq:relation_sre}
    \mathcal{M}_{a}(|\psi\rangle) \leq \frac{a}{a-1} \mathcal{S}_{b}(|\psi\rangle), \quad \text{for } a > 1,\; b \leq 2.
\end{equation}
To prove this, we note that the PE for $k=2$, for states with fixed $S^z = \sum_{j=1}^L Z_j$, can be rewritten as
\begin{equation} \label{eq:s2_pz}
\mathcal{S}_2(|\psi\rangle) = -\log[\zeta^z_2],\quad \zeta_2^z=\sum_{P \in \mathcal{Z}_{L}} \frac{|\langle \psi|P|\psi\rangle|^2}{D},
\end{equation}
where $\mathcal{Z}_{L}\subset \mathcal{P}_L$ denotes Pauli strings made only of $Z$ and $I$ operators. It follows that $\zeta_k \ge \zeta_k^{z}$, and by Jensen's inequality one obtains $\zeta_k^z \ge (\zeta_2^{z})^k$, leading to~\eqref{eq:relation_sre} using monotonicity of Rényi entropy.

A second, sharper inequality is
\begin{equation} \label{eq:relation_sre_2}
\mathcal{M}_{a}(|\psi\rangle) \leq \mathcal{S}_{1/2}(|\psi\rangle), \quad \text{for } a \geq 1/2.
\end{equation}
This follows by bounding the stabilizer norm $\mathcal{D}(\rho) = D^{-1} \sum_{P \in \mathcal{P}_L} |\Tr[P \rho]|$~\cite{howard2017robustness,haug2025efficientwitnessingtestingmagic} in terms of the $l_1$-coherence $C_{l_1} =  \sum_{x \ne x'} |\bra{x}\rho\ket{x'}|$~\cite{baumgratz2014quantifying}, as $\mathcal{D}(\rho) \le 1 + C_{l_1}(\rho)$ (see Supplemental Material~\cite{supmat}). Restricting to pure states and combining with Rényi monotonicity gives~\eqref{eq:relation_sre_2}. 
Taken together, these results formalize the heuristic expectation that, in symmetry-constrained systems, nonstabilizerness and anticoncentration are dynamically and operationally equivalent indicators of quantum complexity.

\paragraph{Pauli and collision matrix-product states.} 
We employ two complementary numerical strategies to compute the PE and SRE in large-scale quantum spin chains:  tensor network methods based on replica matrix-product states (MPS) for scalable computations, and perfect sampling. 

To compute the SRE, we use the perfect sampling~\cite{lami2023nonstabilizernessviaperfect} and the \emph{Pauli-MPS} method introduced in Ref.~\cite{tarabunga2024mps}. The latter approach constructs an MPS $|P(\psi)\rangle$ that encodes the full Pauli spectrum of a state $|\psi\rangle$, with components $\langle \alpha |P(\psi)\rangle = \langle \psi | P_\alpha | \psi \rangle / \sqrt{2^L}$ for Pauli strings $P_\alpha$. Repeated application of a diagonal MPO $W$ yields the replicated vector $|P^{(n)}(\psi)\rangle$, from which the stabilizer Rényi entropy is extracted via its norm. This method maps the SRE to a two-dimensional tensor network that is efficiently contractible using standard MPS routines.

The participation entropy is computed using a \emph{collision-MPS} algorithm introduced in this work (see the End Matter for details). Starting from an MPS representation of $|\psi\rangle$, we construct a diagonal MPO $V$ that encodes the amplitudes of the state in the computational basis. Repeated application of $V$ produces a replicated state $|\psi^{(k)}\rangle$ with components $|\langle x|\psi\rangle|^{2k}$, whose norm yields the PE as $\mathcal{S}_k = (1-k)^{-1} \log \left( \langle \psi^{(k)} | \psi^{(k)} \rangle \right)$. Despite the exponential growth of bond dimension with $k$, iterative compression after each step allows for tractable simulations even in large systems. This framework enables the computation of PE in many-body dynamics at scales that are challenging for conventional Monte-Carlo methods due to sign problems~\cite{luitz2014improving,luitz2014shannon,luitz2014universal,Luitz17}, or based on state-vector simulations due to the exponential growth of the Hilbert space.

\paragraph{Numerical results: Magic resources and transport.} We now present the numerical evidence demonstrating the relationship between stabilizer Rényi entropy and participation entropy in $\mathrm{U}(1)$ quantum many-body systems. Building on this connection, we show that the growth of magic resources offers insights into the underlying quantum transport associated with the $\mathrm{U}(1)$ charge. To this end, we consider a one-dimensional XXZ spin-$1/2$ chain, with and without disorder, of length $L$ as a canonical model to study quantum transport under symmetry constraints. The Hamiltonian reads
\begin{equation}
H = \sum_{i=1}^{L-1} J\left( S_i^x S_{i+1}^x + S_i^y S_{i+1}^y \right) + J_z S_i^z S_{i+1}^z + \sum_{i=1}^{L} h_i S^z_i,
\end{equation}
where $S_i^{\alpha}$ ($\alpha=x,y,z$) are spin-$1/2$ operators on site $i$, and $J_z$ controls the anisotropy and $h_i$ is the random disorder taken uniformly from $h_i \in [-h, h]$. This model conserves total magnetization $S^z = \sum_i S^z_i$ and thus realizes $\mathrm{U}(1)$--symmetric dynamics. Throughout our analysis, we set $J=1$ to fix the energy scale.  

In the clean limit $h=0$, the model is integrable and features different transport regimes depending on the anisotropy parameter $J_z$, while for the disordered case $h\neq 0$, the disorder strength controls different transport regimes. For our analysis, we consider a weak disorder ($  h\lesssim 0.5$) that breaks the integrability and drives the system into a diffusive transport regime~\cite{PhysRevLett.117.040601, Karahalios09, BarLev15, Steinigeweg16, Barisic16}. One approach to distinguishing different transport regimes is to examine the time evolution of spin polarization, starting from an inhomogeneous initial state with opposite magnetization on the two halves of the chain~\cite{ljubotina2017spin}. The net transfer of polarization that captures the melting of the domain wall between the two halves is quantified by $\Delta P(t) = 2\left[P^L(t) - P^R(t)\right]$, where $P^{L/R}(t) = \sum_{i \in L/R} \frac{\langle S_i^z(t) \rangle - \langle S_i^z(0) \rangle}{2}$ are the changes of magnetization on the left and right sides, respectively. The polarization transfer exhibits a power-law growth $\Delta P(t) \sim t^{1/z}$, with the dynamical exponent $z$ characterizing the transport universality class. For the XXZ model, one observes ballistic transport ($z=1$) for $J_z<1$, diffusive behavior ($z=2$) for $J_z>1$, and super-diffusive scaling ($z=3/2$) at the isotropic point $J_z=1$, which is related to the  KPZ universality class~\cite{ljubotina2017spin,PhysRevLett.122.210602,wei2022quantum,ye_universal2022}. 

Here, we study the dynamics of the SRE and PE starting from a domain wall initial state. Fig.~\ref{fig:PE_SRE_ballistic_XXZ} (a)-(b) shows that at weak anisotropy ($J_z=0.25$), both PE and SRE grow linearly in time, $\propto t$, signaling ballistic spreading of quantum complexity. Similarly, Fig.~\ref{fig:PE_SRE_ballistic_XXZ} (c)-(d) illustrates that at $J_z=1$, both entropies grow as $t^\beta$ with $\beta=0.66(3)$, consistent within error bars with the KPZ exponent $z=3/2$ for the time scales considered. The insets in (Fig.~\ref{fig:PE_SRE_ballistic_XXZ}(b, d)) provide the finite-size scaling of the obtained dynamical exponent. In contrast, the entanglement entropy plotted in the insets (Fig.~\ref{fig:PE_SRE_ballistic_XXZ}(a, c)) fails to provide information about the dynamical exponents.  It is worth mentioning that for the domain wall initial state, at the isotropic point, the superdiffusion obtained from the polarization transfer eventually crosses over to diffusion at late time~\cite{Misguich17,wei2022quantum}, and the true superdiffusion appears for the ensemble of high-temperature initial states~\cite{ljubotina2017spin}. Nevertheless, for the considered system sizes and time scales, we do not see a diffusive regime.

\begin{figure}[t]
   \centering
\includegraphics[width=0.45\linewidth]{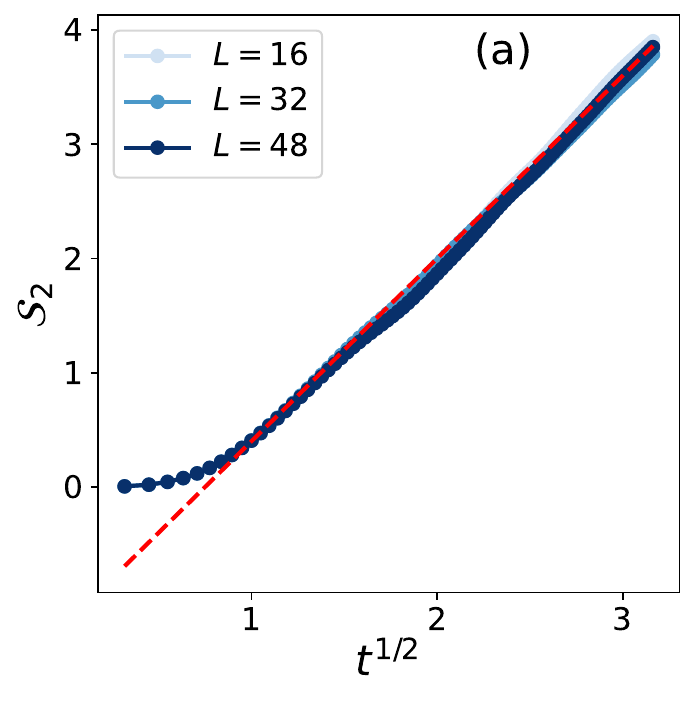}
\includegraphics[width=0.45\linewidth]{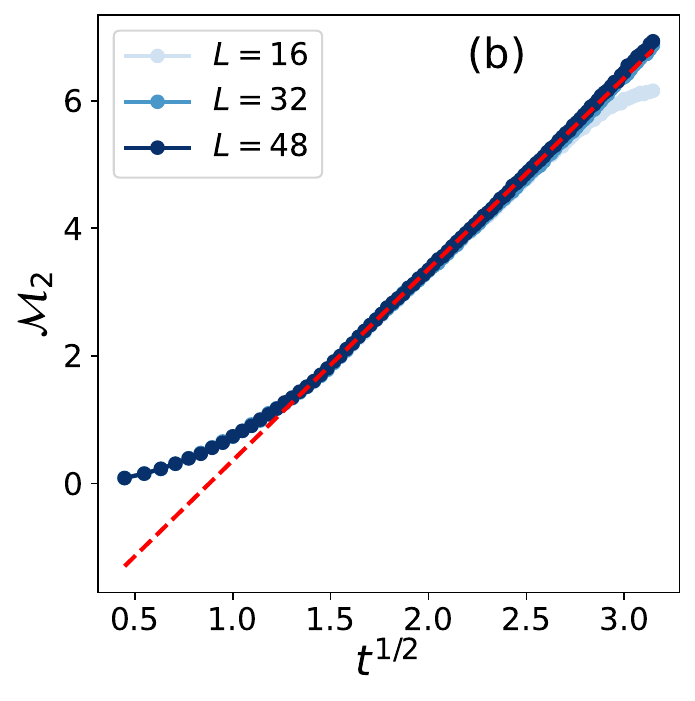}
\caption{\textbf{Diffusive regime in disordered XXZ model ($h=0.2$):} Time evolution of (a) PE and (b) SRE for different system sizes $L$. Both complexity measures exhibit clear diffusive scaling, growing as $t^{1/2}$. This diffusive behavior emerges due to the introduction of weak disorder, which breaks integrability and enforces a diffusive transport regime. }
\label{fig:diffusive_regime_XXZ_model}
\end{figure}

Next, we considered the disordered case, and fixed the disorder strength to $h=0.2$. We use time-evolution using TDVP up to $L=48$ sites and perform a disorder averaging over an ensemble of $40$ different realizations. In Fig.~\ref{fig:diffusive_regime_XXZ_model}, we show the evolution of PE and SRE, and as can be seen, both quantities grow as a power law $t^{1/2}$, in agreement with diffusive transport. 

\paragraph{Discussion.}
We have investigated the interplay between transport and quantum complexity in 
$U(1)$-symmetric systems through the lens of two key resource-theoretic diagnostics: the SRE, capturing nonstabilizerness, and the PE, capturing anticoncentration. Our results demonstrate that both measures exhibit universal power-law growth governed by the transport dynamics—ballistic, diffusive, or superdiffusive—across clean and disordered XXZ spin chains. This reveals a deep connection between transport, usually seen as an emergent classical effect, and the complexity growth of many-body states under symmetry-constrained unitary dynamics. Moreover, we uncover a direct link between the time evolution of two a priori distinct resources—anticoncentration and nonstabilizerness—in U(1)-symmetric many-body dynamics.
We supported our claims analytically by linking PE and SRE through formal bounds, and numerically via a scalable collision MPS algorithm that accurately computes PE in large-scale dynamics.

Our findings suggest several directions for future work, including extending this framework to other conservation laws and exploring non-Abelian or Floquet dynamics~\cite{moessner2017equilibration}. The connection between nonstabilizerness or anticoncentration and hydrodynamic modes also warrants further investigation~\cite{kheman,PhysRevX.8.021013,tibor2019sub,Michailidis24Corrections}. 
Finally, our predictions are testable on current experimental platforms such as trapped ions~\cite{joshi2022observing} and cold atoms~\cite{wei2022quantum}, which realize $U(1)$-symmetric dynamics. In these setups, the participation entropy can be obtained from the wavefunction snapshots either by building the probability distribution or by performing principal component analysis (PCA) on the wavefunction snapshot dataset and computing the PCA entropy~\cite{bhakuni2024diagnosing}.

\begin{acknowledgments}
\paragraph{Acknowledgements.}
We thank M. Collura and J. Goold for inspiring discussions. 
E.\,T. acknowledges support from  ERC under grant agreement n.101053159 (RAVE), and
CINECA (Consorzio Interuniversitario per il Calcolo Automatico) award, under the ISCRA 
initiative and Leonardo early access program, for the availability of high-performance computing resources and support.
P.S.T. acknowledges funding by the Deutsche Forschungsgemeinschaft (DFG, German Research Foundation) under Germany’s Excellence Strategy – EXC-2111 – 390814868.
M.\,D. was partly supported by the QUANTERA DYNAMITE PCI2022-132919, by the EU-Flagship programme Pasquans2, by the PNRR MUR project PE0000023-NQSTI, the PRIN programme (project CoQuS), and the ERC Consolidator grant WaveNets.
X.T. acknowledges support from DFG under Germany's Excellence Strategy – Cluster of Excellence Matter and Light for Quantum Computing (ML4Q) EXC 2004/1 – 390534769, and DFG Collaborative Research Center (CRC) 183 Project No. 277101999 - project B01.
P.S. acknowledges fellowship within the “Generación D” initiative, Red.es, Ministerio para la Transformación Digital y de la Función Pública, for talent atraction (C005/24-ED CV1), funded by the European Union NextGenerationEU funds, through PRTR.
\end{acknowledgments}

\textit{Note Added. ---}
During the completion of this work, we became aware of a complementary study by ~\cite{sticlet2025non}, which discusses nonstabilizerness in open spin chains. While the two works are different in all technical aspects, both support the relation between nonstabilizerness and universality classes of transport.

\clearpage

\begin{figure*}[t]
   \centering
\includegraphics[width=1.\linewidth]{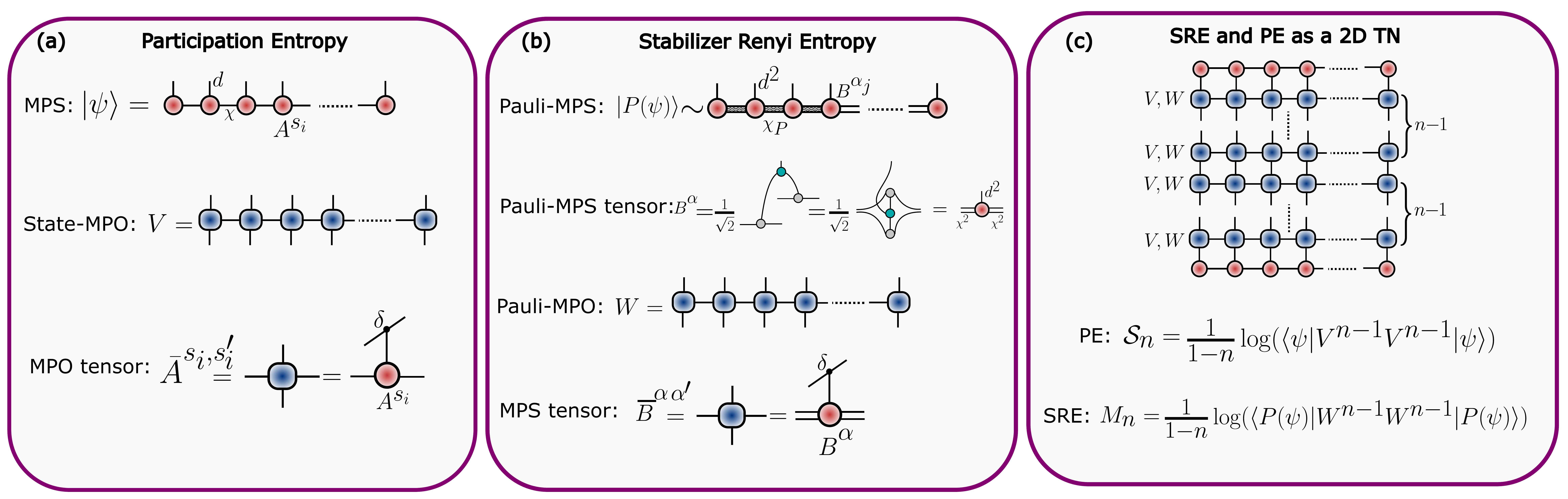}
\caption{\textbf{Sketch of Replica TN for PE and SRE:} 
(a) Definitions of tensors used to construct the MPO to calculate the PE.
(b) Definitions of tensors used for the construction of Pauli-MPS. In particular, we show how to construct the tensors of the Pauli-MPS and MPO $B$ and $B^{\alpha \alpha^{\prime}}$. (c) The PE and SRE are represented as contractions of a two-dimensional tensor network. }
\label{fig:sketch_PE_SRE_2D_TN}
\end{figure*}

\section{End Matter.}
To compute the stabilizer Rényi entropy (SRE) and participation entropy (PE), we present two complementary approaches. First, we introduce a \textit{collision MPS} algorithm based on replica tensor network contraction, which enables efficient evaluation of the PE for time-evolved quantum states represented as matrix product states. This method extends the framework developed in Ref.~\cite{tarabunga2024mps}. More concretely, we represent the collision and the Pauli MPS in the Fig.~\ref{fig:sketch_PE_SRE_2D_TN} (a)-(b). For both  the SRE and PE, we
express them as a two-dimensional tensor network as shown in Fig.~\ref{fig:sketch_PE_SRE_2D_TN} (c), thereby enabling approximate contraction
using established MPS methods.

Second, we discuss the application of perfect sampling techniques for computing PE~\cite{PhysRevB.85.165146,stoudenmire2010minimally,lami2023nonstabilizernessviaperfect},
which allow unbiased estimation of participation quantities by sampling amplitude distributions in the computational basis. These tools provide robust and scalable access to PE and SRE in large-scale many-body simulations.

\textit{Collision and Pauli MPS:}\label{sec:met_collision_pauli}
We numerically calculate the full-state SRE and the PE using the replica MPS.
In particular, to calculate the SRE we used the replica Pauli MPS method developed in Ref.~\cite{tarabunga2024mps}. This method begins with an  MPS representation of the state $|\psi\rangle$.  
In Fig.~\ref{fig:sketch_PE_SRE_2D_TN}, we present the algorithms to compute the PE and the SRE through a 2D TN. In particular in Fig.~\ref{fig:sketch_PE_SRE_2D_TN} (a), we show how to construct the Pauli-MPS, and in (b) we show the definition of the tensors used to compute the PE. This allows one to compute not only the SRE, but also the PE.
For  both the SRE and PE, we express it as a two-dimensional tensor network as shown in Fig.~\ref{fig:sketch_PE_SRE_2D_TN}(c), thereby enabling approximate contraction using established MPS methods.

Let us consider a system of $L$ qubits in a pure state $|\psi \rangle$ given by an MPS of bond dimension $\chi$:
\begin{equation} \label{eq:mps}
    |\psi \rangle=\sum_{s_1,s_2,\cdots,s_L} A^{s_1}_1 A^{s_2}_2 \cdots A^{s_L}_L |s_1,s_2,\cdots s_L \rangle
\end{equation}
with $A_i^{s_i}$ being $\chi$\,$\times$\,$\chi$ matrices, except at the left (right)
boundary where $A_1^{s_1}$ ($A_L^{s_L}$) is a $1$\,$\times$\,$\chi$ ($\chi$\,$\times$\,$1$) row (column) vector. Here $s_i$\,$\in$\,$\left \lbrace 0, 1 \right \rbrace$  is a local computational basis.
The state is assumed right-normalised, namely $\sum_{s_i} A_i^{s_i \dagger} A_i^{s_i}$\,$=$\,$1$. 
To calculate the PE, we define a diagonal operator $V$ whose diagonal elements are components of the MPS vector $|\psi\rangle$, $\langle \boldsymbol{s^{\prime}}| V |\boldsymbol{s} \rangle = \delta_{\boldsymbol{s^{\prime}},\boldsymbol{s}} \langle \boldsymbol{s^{\prime}} |\psi\rangle$. The MPO form of the $V$ reads
\be 
V= \sum_{\boldsymbol{s},\boldsymbol{s^{\prime}}} \overline{A}^{s_1,s^{\prime}_1}_1 \overline{A}^{s_2,s^{\prime}_2}_2 \cdots \overline{A}^{s_N,s^{\prime}_N}_N |s_1, \cdots, s_N \rangle \langle s^{\prime}_1, \cdots, s^{\prime}_N |
\ee
where $\overline{A}^{s_i,s^{\prime}_i}_i=\delta_{s_i,s^{\prime}_i} A^{s_i}$. Applying the MPO $V$ $k-1$ time to $|\psi\rangle$, we obtain $|\psi^{(k)}\rangle=V^{k-1}|\psi\rangle$, which is a vector with element $\langle \boldsymbol{s}|\psi^{(k)}\rangle=p^k_{\boldsymbol{s}}=\left|A^{s_1} \cdots A^{s_L} \right|^k$.  
We denote the tensor of $|\psi^{(k)}\rangle$ by $A^{(k)s_i}_i$, and we have
\be
\sum_{s} p^k_s=\langle \psi^{(k)} | \psi^{(k)} \rangle \Rightarrow \mathcal{S}_k=\frac{1}{1-k} \log\left( \langle \psi^{(k)} | \psi^{(k)} \rangle \right) .
\ee
The exact bond dimension of$|\psi^{(k)} \rangle$ is the min between $(\chi^{k}, 2^{L/2}) $ 
i.e., for large systems it grows exponentially with the order $k$.  However, by interpreting it as the repeated application of an MPO $V$ onto an MPS, we can sequentially compress the resulting MPS
after every iteration, and keep the best description of the resulting state as an MPS with some upper-bounded bond dimension $\chi_C$.

\begin{figure}[b!]
   \centering
\includegraphics[width=0.48\linewidth]{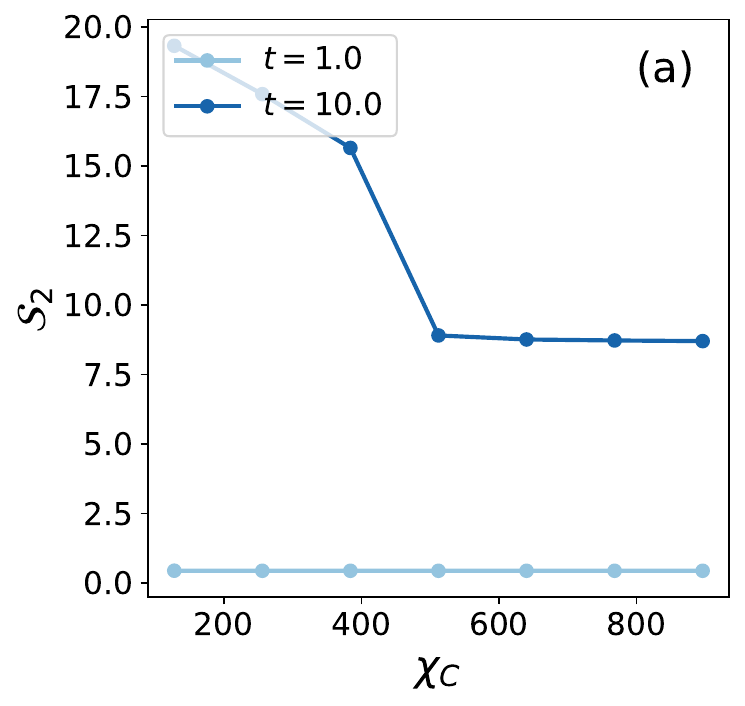}
\includegraphics[width=0.45\linewidth]{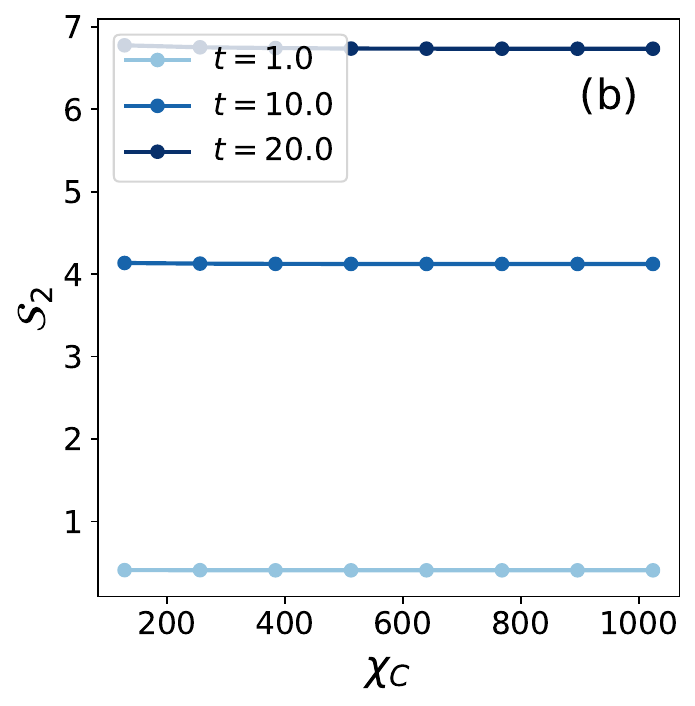}
\caption{\textbf{Convergence of collision MPS for $N=80$:}  (a)Participation entropy $\mathcal{S}_2$ as a function of the bond dimension cutoff $\chi_C$ for different time $t=1$ and $t=10$ in the balistic regime ($J_z=0.25$), (b) $\mathcal{S}_2$ in the superdiffusive regime ($J_z=1.0$)  indicating that convergence is preserved across dynamical evolution.
The saturation of $\mathcal{S}_2$ with increasing $\chi_C$ confirms that the iterative compression strategy in the collision MPS method reliably captures the participation entropy even in large-scale, time-evolved states.}
\label{fig:convergence}
\end{figure}

To validate the accuracy and scalability of the collision MPS method, we benchmarked the participation entropy as a function of the bond dimension $\chi_C$. As shown in the Fig.~\ref{fig:convergence}, the computed values of $\mathcal{S}_2$ converge rapidly with increasing $\chi_C$ both at short and long times. This confirms that the sequential MPO application and iterative compression used in the collision MPS scheme provide an efficient and controlled approximation of the replicated state $|\psi^{(k)\rangle}$ even as its formal bond dimension grows exponentially with $k$. The method thus enables accurate estimation of PE in time-evolved quantum states far beyond the reach of exact state-vector simulations, making it a powerful tool for tracking complexity growth in many-body dynamics.

To calculate the SRE, we used the replica Pauli-MPS. Let us define the binary string $\boldsymbol{\alpha}$\,$=$\,$(\alpha_1, \cdots, \alpha_N)$ with $\alpha_j$\,$\in$\,$\{00,01,10,11\}$.
The Pauli strings are defined as $P_{\boldsymbol{\alpha}} $\,$=$\,$ P_{\alpha_1}  \otimes P_{\alpha_2} \otimes \cdots  \otimes P_{\alpha_L}$
where $P_{00}$\,$=$\,$I, P_{01}$\,$=$\,$\sigma^x, P_{11}$\,$=$\,$\sigma^y,$ and $P_{10}$\,$=$\,$\sigma^z$. We define the Pauli vector of $|\psi \rangle$ as $|P(\psi) \rangle$ with elements $  \langle \boldsymbol{\alpha} |P(\psi) \rangle$\,$=$\,$ \langle \psi | P_{\boldsymbol{\alpha}} |\psi \rangle / \sqrt{2^L}$.  When $|\psi \rangle$ has an MPS structure as in Eq. \eqref{eq:mps}, the Pauli vector can also be expressed as an MPS as follows
\begin{equation} \label{eq:PauliMPS}
    |P(\psi) \rangle=\sum_{\alpha_1,\alpha_2,\cdots,\alpha_N} B^{\alpha_1}_1 B^{\alpha_2}_2 \cdots B^{\alpha_L}_L | \alpha_1, \cdots, \alpha_L \rangle
\end{equation}
where $B^{\alpha_i}_i $\,$=$\,$ \sum_{s,s'} \langle s | P_{\alpha_i} | s' \rangle A^s_i \otimes \overline{A^{s'}_i} / \sqrt{2}$ are $\chi^2 \times \chi^2$ matrices. Note that the MPS is normalized due to the relation $\frac{1}{2^N} \sum_{\boldsymbol{\alpha}} \langle \psi | P_{\boldsymbol{\alpha}} | \psi \rangle^2 $\,$=$\,$ 1$ which holds for pure states.
To calculate the SREs, we define a diagonal operator $W$ whose diagonal elements are the components of the Pauli vector, $\langle  \boldsymbol{\alpha'}| W |\boldsymbol{\alpha} \rangle= \delta_{\boldsymbol{\alpha'},\boldsymbol{\alpha}} \langle \boldsymbol{\alpha'} |P(\psi) \rangle$. The MPO form of $W$ reads
\begin{equation}
    W =\sum_{\boldsymbol{\alpha},\boldsymbol{\alpha'}} \overline{B}^{\alpha_1,\alpha'_1}_1 \overline{B}^{\alpha_2,\alpha'_2}_2 \cdots \overline{B}^{\alpha_L,\alpha'_L}_L | \alpha_1, \cdots, \alpha_N \rangle \langle \alpha'_1, \cdots, \alpha'_N |
\end{equation}
where $\overline{B}^{\alpha_i,\alpha'_i}_i$\,$=$\,$ B^{\alpha_i}_i \delta_{\alpha_i,\alpha'_i}$.
Applying $W$ $n$\,$-$\,$1$ times to $|P(\psi) \rangle$, we obtain $|P^{(n)}(\psi) \rangle $\,$=$\,$W^{n-1} |P(\psi) \rangle$, which is a vector with elements $\langle \boldsymbol{\alpha} |P^{(n)}(\psi) \rangle$\,$=$\,$\langle \psi | P_{\boldsymbol{\alpha}} |\psi \rangle^n / \sqrt{2^{Ln}}$. 
We denote the local tensors of $|P^{(n)}(\psi) \rangle$ by $B^{(n)\alpha_i}_i$.
We have $\sum_{\boldsymbol{\alpha}} \langle \psi | P_{\boldsymbol{\alpha}} | \psi \rangle^{2n}/2^{Nn} = \langle P^{(n)}(\psi)  | P^{(n)}(\psi)  \rangle$ 
and 
\begin{equation} \label{eq:sre_replica}
    M_n = \frac{1}{1-n} \log{\langle P^{(n)}(\psi)  | P^{(n)}(\psi)  \rangle} - L\log2.
\end{equation}
The Pauli-MPS itself can also be approximated with a bond dimension $\chi_P$\,$<$\,$\chi^2$. The computational cost of this compression is $O\left( \chi_P^2 \chi^2 + \chi^3 \chi_P \right)$. This is particularly advantageous for states with exponentially decaying entanglement spectrum (e.g., gapped phases), in which case $\chi_P$ can be truncated to a value much smaller than $\chi^2$.

\textit{Perfect sampling} \label{sec:met_perfect_sempling}
Computing the PE and the SRE requires the evaluation of the expectation value of the generic power of $\Pi^{n-1}_{\rho}$ (or $\log(\Pi_{\rho})$ for $n=1$)  over the probability distribution $\Pi_{\rho}$ itself.  In the case of PEs $\Pi_{\rho}=p_{\boldsymbol{x}}$ instead in the case of SREs $\Pi_{\rho}=\Xi_{\rho}$.
This suggests a natural way to estimate the PEs and SREs,
based on a sampling from $\Pi_{\rho}$.

As regards the PEs, the task of sampling from the set of bit strings $\boldsymbol{s}$,  which has size $D=2^N$ may appear as exponentially hard. To overcome this difficulty, we rewrite the full probability in terms of conditional and prior (or marginal) probabilities as
\be \label{eq:Prob_sampling_PE}
\Pi(\boldsymbol{s})=\Pi(s_1) \Pi(s_2|s_1) \cdots \Pi(s_N|s_1 \cdots s_N) 
\ee
where $\Pi(s_j|s_1 \cdots s_{j-1})$ is the probability that the Pauli matrix to select $s_j$ at the position $j$ give that the bitstring $s_1, \cdots, s_{j-1}$ has already occurred at the position $1 \cdots j-1$, no matter the occurrences in the rest of the system $1,\cdots,j-1$, no matter the occurrences in the rest of the system. Following the conditional sampling prescription described in the previous section, we start from the first term of the expansion in Eq.~(\ref{eq:Prob_sampling_PE}). This can be written as
\be
\Pi(s_1)=\sum_{s_2,\cdots,s_N} \langle s_1 s_2 \cdots s_N | \psi \rangle \langle s_1 s_2 \cdots s_L | \psi \rangle^{*}.  
\ee
Using the structure of the MPS and the fact that it is in right canonical form,  we can express the previous probability as follows
\be 
\Pi(s_1) 
=A^{s_1} A^{s_1, *} 
\ee
After evaluating $\Pi(s_1)$, one can extract a sample from this distribution, obtaining the first element of the string. The calculation of the next terms of Eq.(\ref{eq:Prob_sampling_PE}) and the
extraction of the remaining terms proceeds following the
same line. The full sampling recipe is summarized in Fig. \ref{fig:sketch_PE_SRE_2D_TN}.  If we sample
according to the probablity $\Pi(\boldsymbol{s})$, $\mathcal{S}_k$ can be estimated using the unbiased
estimators
\be 
\mathcal{S}_k=\frac{1}{1-k} \frac{1}{\mathcal{N}} \sum_{\mu}  \Pi_\rho \left(\boldsymbol{s}_{\mu} \right)^{k-1}
\ee
for $k>1$, while for $k=1$,
\be 
\mathcal{S}_1=\frac{1}{\mathcal{N}} \sum_{\mu} \log \Pi_\rho \left(\boldsymbol{s}_{\mu} \right)\;.
\ee

\bibliography{ref}
\newpage 
\cleardoublepage
\setcounter{section}{0}
\setcounter{page}{1}
\setcounter{equation}{0}
\thispagestyle{empty}
\begin{center}
\textbf{\large Supplemental Material for ``Universal Spreading of Nonstabilizerness and Quantum Transport''}\\

\end{center}

\twocolumngrid
\pagenumbering{roman}

\setcounter{equation}{0}
\setcounter{figure}{0}
\renewcommand{\theequation}{S\arabic{equation}}
\renewcommand{\thefigure}{S\arabic{figure}}
\renewcommand{\thetable}{S\arabic{table}}

\section{Relation between PE and SRE }
We provide here the analytical derivation of the inequalities connecting stabilizer Rényi entropy (SRE) and participation entropy (PE) in the presence of U(1) symmetry.
In fact, it can be shown that the PE is related to the SRE by the following inequality:
\begin{equation} \label{eq:relation_sre}
    M_{a}(|\psi \rangle) \leq \frac{a}{a-1} \mathcal{S}_{b}(|\psi \rangle) \quad (a >1, b \leq 2).
\end{equation}
To prove this relation, we will first show that
\begin{equation} \label{eq:s2_pz}
    \mathcal{S}_2(|\psi \rangle) = -\log \frac{\sum_{P\in \mathcal{Z}_L} |\langle \psi|P|\psi\rangle|^2}{2^L}
\end{equation}
where $\mathcal{Z}_L$ is the group of Pauli operators which only consist of $I$ and $Z$ operators. To see this, we write
\begin{equation}
    \sum_x |\langle x | \psi \rangle|^{4} = \Tr([ \ket{00}\bra{00}+\ket{11}\bra{11}]^{\otimes L} \ket{\psi} \bra{\psi}^{\otimes 2}).
\end{equation}
One can see that $\ket{00}\bra{00}+\ket{11}\bra{11}=(I\otimes I+Z\otimes Z)/2$, which implies
\begin{equation}
\begin{split}
    \sum_x |\langle x | \psi \rangle|^{4} &= \frac{\Tr([ I\otimes I+Z\otimes Z]^{\otimes L} \ket{\psi} \bra{\psi}^{\otimes 2})}{2^L} \\
    &= \frac{\sum_{P\in \mathcal{Z}_L} ||\langle \psi|P|\psi\rangle|^2}{2^L},
\end{split}
\end{equation}
thus Eq. \eqref{eq:s2_pz} is proven.

Now, we have
\begin{equation*} 
    \begin{split}
        \frac{1}{2^L} \sum_{P\in \mathcal{P}_L} |\langle \psi|P|\psi\rangle|^{2n} &\geq \frac{1}{2^L} \sum_{P\in \mathcal{Z}_L} |\langle \psi|P|\psi\rangle|^{2n} \\
        & \geq  \left( \frac{\sum_{P\in \mathcal{Z}_L} |\langle \psi|P|\psi\rangle|^{2}}{2^L} \right)^{n} \\
        & =  \left( \sum_{x} |\langle x | \psi \rangle|^{4} \right)^{n} \\
    \end{split}
\end{equation*}
Taking logarithm on both sides and dividing by $(n-1)$, we arrive at Eq.~\eqref{eq:relation_sre} for $b=2$. The final inequality is obtained by using the hierarchy of R\'enyi entropies $S_a \leq S_b$ for $a \geq b$.

One can also show the following inequality:
\begin{equation} \label{eq:relation_sre_2}
    M_{a}(|\psi \rangle) \leq  \mathcal{S}_{1/2}(|\psi \rangle) \quad (a \geq 1/2).
\end{equation}
To show this, we will prove the inequality for mixed states $\rho$:
\begin{equation} \label{eq:d_c}
    \mathcal{D}(\rho) \leq 1 + C_{l_1}(\rho),
\end{equation}
where $\mathcal{D}(\rho)=
2^{-N}\sum_{P\in\mathcal{P}_N} \lvert \text{tr}[ P \rho] \rvert$ is the stabilizer norm and $C_{l_1}(\rho)=
\sum_{x \neq x'} \lvert  \bra{x} \rho \ket{x'}  \rvert$ is the $l_1$ norm coherence. Note that the stabilizer norm is a witness of magic for mixed states~\cite{howard2017robustness,haug2025efficientwitnessingtestingmagic} and the $l_1$ norm coherence is a coherence monotone~\cite{baumgratz2014quantifying}. Restricting to pure states leads directly to Eq.~\eqref{eq:relation_sre_2} for $a=1/2$, and the final inequality is again obtained by the hierarchy of R\'enyi entropies.

We label the Pauli strings using a pair of indices $a,a'\in \{ 0,1\}$, such that $P_{a,a'} = i^{aa'} X^a Z^{a'}$ whereby $P_{0,0} = I$, $P_{1,0} = X$, $P_{0,1} = Z$, and $P_{1,1} = Y$. We have
\begin{equation}
\begin{split}
\mathcal{D}(\rho) 
&= 
\frac{1}{D} \sum_{\mathbf{a},\mathbf{a'}} 
\left\vert 
 \Tr(\rho P_{\mathbf{a},\mathbf{a'}})  
\right\vert \\
&= 
\frac{1}{D} \sum_{\mathbf{a},\mathbf{a'}} 
\left\vert 
\sum_{x}  \langle x | \rho P_{\mathbf{a},\mathbf{a'}} | x \rangle 
\right\vert \\
&= \frac{1}{D} \sum_{\mathbf{a},\mathbf{a'}} 
\left\vert
\sum_{x}  
\prod_{j=1}^N 
i^{a_j a'_j} \prod_{j=1}^N 
 (-1)^{x_j a'_j} \langle x_j | \rho | a_j \oplus x_j \rangle \right\vert\\
&\leq \frac{1}{D } \sum_{\mathbf{a},\mathbf{a'}} 
\sum_{x}\left\vert
 \langle x | \rho | a \oplus x \rangle 
\right\vert \\
&= \sum_{\mathbf{a}} 
\sum_{x}\left\vert \langle x | \rho | a \oplus x \rangle 
\right\vert \\
&= 1 + \sum_{ x\neq x'}\left\vert \langle x | \rho | x' \rangle \right\vert\\
&= 1 + C_{l_1}(\rho)\, . 
\end{split}
\end{equation}

\end{document}